\documentstyle[overcite,eqsecnum,preprint,aps,prd]{revtex} 
\input epsf
\tighten
\begin{document}
\draft

\title{ The Imprint of Gravitational Waves in Models Dominated by 
a  Dynamical Cosmic Scalar Field }

\author{ 
R. R. Caldwell 
and 
Paul J. Steinhardt
}

\address{Department of Physics and Astronomy\\
University of Pennsylvania\\
Philadelphia, PA 19104}

\maketitle

\begin{abstract}

An alternative to the standard cold dark matter model has been recently proposed
in which a significant fraction of the energy density of the universe is due to a
dynamical scalar field ($Q$) whose effective equation-of-state differs from that
of matter, radiation or cosmological constant ($\Lambda$).  In this paper, we
determine how the $Q$-component modifies the primordial inflation gravitational
wave (tensor metric) contribution to the cosmic microwave background anisotropy
and, thereby, one of the key tests of inflation.

\end{abstract} 

\section{Introduction}
\label{sectionintro}

A key prediction of inflationary cosmology is a nearly scale-invariant spectrum
of initial fluctuations composed of energy density (scalar
metric)\cite{GP,Hawking,Starobinskii,BST} and gravitational wave (tensor
metric)\cite{Rub,Staro,AW} perturbations. The two components leave imprints on
the cosmic microwave background (CMB) anisotropy that are potentially
detectable and distinguishable.  The spectrum of each component is
characterized by a spectral index ($n$) which determines how the perturbation
amplitude varies with wavelength.  Each spectrum is also predicted to be nearly
scale-invariant;  using the standard notational convention, this corresponds to
$n_S \approx 1$ for the scalar metric fluctuations  and $n_T \approx 0$ for the
tensor metric fluctuations.  The spectral indices and the ratio of tensor to
scalar amplitudes on any given scale are determined by the equation-of-state
during inflation or, equivalently, the inflaton potential.  A prediction of
inflation is a simple relation between $n_S$, $n_T$ and the ratio of tensor to
scalar amplitudes that applies to nearly all inflationary
models.\cite{Davisetal} Unlike other features of inflation, such as flatness
or a nearly Harrison-Zel'dovich spectra, the predicted relation between the
spectral indices and amplitudes was unanticipated prior to the development of
the inflationary model and is, in this sense, a unique stamp of inflation.

It is hoped that forthcoming measurements of CMB
anisotropy\cite{KnoxTurner,Jungman}, complemented by measurements of CMB
polarization\cite{CMBgravTest}, can be used to detect a tensor component and to
test the inflationary scenario. The CMB temperature anisotropy power spectrum
can be expressed in multipole moments $C_{\ell}$, where each $C_{\ell}$ can be
divided into a sum of independent  tensor ($C_{\ell}^{(T)}$) and scalar
($C_{\ell}^{(S)}$) subcomponents. While the individual tensor and scalar 
contribution to the temperature anisotropy cannot be observed independently,
the presence of a strong tensor component may be measured through correlations
of polarization and temperature.\cite{Selj,Kamion,Sperg} The amplitude of these
correlations, characterized by the ratio of  tensor-to-scalar multipole
moments,  $r_{\ell} \equiv C_{\ell}^{(T)}/C_{\ell}^{(S)}$, can be used to test 
the inflationary scenario.  However, the moments do not depend on the
primordial relation between spectral indices and amplitudes alone. The power
spectrum also depends on the evolutionary history since inflation ended.  For
example, in the standard cold dark matter (sCDM) model  in which $\Omega_m =1$,
the inflationary prediction\cite{Davisetal,Oth} is well-known to be
$$
r_2 \approx 7(1-n_S).
$$
This prediction is modified  for open CDM models or CDM models with a
cosmological
constant ($\Lambda$), in which the post-inflation evolution is different, as has
been studied previously.\cite{KnoxPRD}

The purpose of  this paper is to discuss  the gravitational wave contribution
to the CMB anisotropy in a new class of cosmological models where a significant
fraction of the energy density of the universe takes the form of a cosmic
scalar field ($Q$) with an equation-of-state different from that of matter,
radiation or cosmological constant.\cite{CDS} The scalar field component of the
cosmic fluid has been dubbed ``quintessence" and cosmological models based on a
combination of quintessence and cold dark matter components are known as QCDM
models. In a recent paper\cite{CDS} (henceforth, referred to as Paper I), we
computed  the background evolution, the CMB power spectrum and the mass
spectrum for QCDM models. We showed that models of this type result in  a
significantly better fit than sCDM to the CMB temperature power spectrum, the
mass power spectrum, early structure formation, and distant supernovae and
gravitational lens count measurements. In Paper I we restricted our attention
to the case of strictly scale-invariant ($n_S=1$) spectra of purely scalar
metric perturbations. In the present paper, we discuss how the tensor
contribution to the CMB anisotropy is modified by the presence of a
$Q$-component. We compute the full, scalar plus tensor CMB power spectrum for
QCDM models based on inflationary initial conditions and determine how the
relation between primordial spectral indices and amplitudes predicted by
inflation is modified in QCDM models. The result is a generalization of a key
test of inflationary cosmology.

The organization of this paper is as follows. In Section \ref{section2}  we
briefly review the predictions of inflation for the spectrum of initial tensor
and scalar perturbations and the resultant contributions to the CMB
anisotropy.  In Section \ref{section3} we discuss how the computation of the
tensor contribution is modified in QCDM models. (The computation of the scalar
spectrum has been described in Paper I.)  We then present the computed  tensor
contribution to the CMB power spectrum for QCDM models 
and compare with an sCDM model.
In Section \ref{section4}  we present our key result, the generalization of the
inflationary relation between spectral index and amplitude for QCDM models.

\section{Review of Inflationary Predictions for the sCDM Model}
\label{section2}

Inflation predicts a  nearly scale-invariant power  spectrum of energy density
(scalar metric)  and gravitational wave (tensor metric) fluctuations.  For both
subcomponents, the power spectrum as a function of Fourier mode $k$ can be
expressed to lowest order in terms of the Hubble constant and its
time-derivatives evaluated when the Fourier mode $k$ was stretched beyond the
horizon during inflation, at $k=aH$. Here $a$ is the Friedmann-Robertson-Walker
scale factor and $H=H(\phi)$ is the Hubble parameter, which depends on the
expectation value of the inflaton field, $\phi$.  Expanding around some given
wavenumber $k_0$, the power spectrum can be parameterized by spectral
amplitudes $A_{(S,T)}$ and spectral indices $n_{(S,T)}$:
\begin{eqnarray}
P_S(k) &=& A_S^2  \left(\frac{k}{k_0}\right)^{n_S -1}  
= 4 \left(\frac{H^2}{m_p^2 |H'|} \right)^2 {\large|}_{k=aH} 
\\ 
P_T(k) &=& A_T^2  \left(\frac{k}{k_0}\right)^{n_T}  =  
\frac{16}{\pi} \left( \frac{H}{m_p}
\right)^2 {\large |}_{k=aH},
\end{eqnarray}
where $H' = dH/d\phi$ and $m_p$ is the Planck mass.

A signature of inflation is the series of relations between spectral indices
and amplitudes.\cite{Davisetal}   The tensor spectral index,   $H(\phi)$ and
the power spectrum amplitudes obey the relation\cite{Davisetal,Oth}
\begin{equation}
n_T  =  - \frac{m_p^2}{2 \pi} \left( \frac{H'}{H}\right)^2
= -\frac{1}{8} \frac{A_T^2 }{A_S^2 }.
\end{equation}
The ratio of power spectrum amplitudes can be related to the ratio of
tensor-to-scalar contributions to the CMB anisotropy quadrupole moment
\begin{equation} \label{key1}
r_2 \equiv \frac{C_2^{(T)}}{C_2^{(S)}} \approx - 7 n_T.
\end{equation}
We will comment on the coefficient below.  The scalar spectral index satisfies 
a more complicated relation:
\begin{equation}
n_S -1 = n_T -  \frac{m_p^2}{2 \pi} \,
\left(\frac{H'}{H}\right)'.
\end{equation}
However, for all but an exceptional set of inflaton potentials, the Hubble
parameter evolves so slowly during inflation that the second term is negligible
and $n_S-1 \approx n_T$.   In this case,
\begin{equation} \label{key}
r_2 \approx 7 (1-n_S).
\end{equation}
Hence, the relations arise because all parameters associated with the
perturbation spectrum are ultimately determined only by $H$ and its derivatives
during inflation.

The coefficient in Eqs.~(\ref{key1}) and (\ref{key}) is crucial for the
purposes of this paper. In the sCDM model, the coefficient is set largely by
the Sachs-Wolfe effect, the variation of the gravitational potential on the
surface of last scattering, which is only very weakly dependent on model
parameters such as the Hubble constant or the baryon density. Hence,
Eq.~(\ref{key}) has been presented in the literature\cite{Davisetal,Oth} as a
robust prediction of inflation. However, this prediction is valid {\it only} if
$\Omega_m=1$. In models in which the matter density is less than unity, such as
the QCDM models considered in this paper or models with a cosmological
constant, there is a large, integrated Sachs-Wolfe contribution to the large
angle CMB anisotropy, which changes the predicted relationship between the
spectral indices and the tensor-to-scalar ratio. Because the integrated
Sachs-Wolfe effect depends on the time variation of the gravitational potential
along the line-of-site to the last scattering surface, the amplitude of the
effect is sensitive to the present value of $\Omega_m= 1-\Omega_Q$, the
equation-of-state $w$, and the variation of $w$ with time.  Consequently, the
coefficient in Eq.~(\ref{key}) is modified by a model-dependent function 
(Section \ref{section4}), the central result of this paper.

\section{Tensor Contribution to the CMB Anisotropy in QCDM Models}
\label{section3}

In this section we discuss the QCDM models for which we have computed the
gravitational wave contribution to the CMB anisotropy. We present the equations
necessary to evolve the background and tensor perturbations. We discuss the
properties of the Q-matter 
and their dependence on the effective potential for $Q$, $V(Q)$,
 and identify two broad categories of
models. This classification  simplifies our
survey,  in the following section, of the imprint of gravitational waves in
QCDM models. Finally, we present some sample results of our computation of the
tensor contribution to the CMB anisotropy spectrum.

\subsection{Background Equations}

The QCDM models are constructed from spatially flat, Friedmann-Robertson-Walker
(FRW) space-times containing baryons, cold dark matter, neutrinos, radiation
and a cosmic scalar field or Q-component. The space-time metric is given by
$ds^2 = -dt^2 + a^2(t) d\vec x^2$ where $a$ is the expansion scale  factor and
$t$ is the cosmological time. The background equation of motion, the energy
density, and the pressure for the Q-matter are
\begin{eqnarray} 
&& \ddot Q + 3 H \dot Q = - {\partial V \over \partial Q}, 
\label{Qbackground1} \\
&& \rho_Q = {1 \over 2 }\dot Q^2 + V, \qquad 
p_Q = {1 \over 2}\dot Q^2 - V ,
\label{Qbackground2}
\end{eqnarray}
where the dot represents $\partial/\partial t$. Hence, the relations in
(\ref{Qbackground1}-\ref{Qbackground2}) supplement the usual background
equations  to specify the evolution of all components of the cosmological 
fluid.

\subsection{Gravitational Wave Equations}

Fluctuations about the background space-time can be represented  in the
synchronous gauge, where the linearized space-time metric is
\begin{equation}
ds^2 =  -dt^2 + a^2(t)(\gamma_{ij} + h_{ij})dx^i dx^j.
\end{equation}
Here, $\gamma_{ij}$ is the unperturbed spatial metric and  $h_{ij} = h^S_{ij} +
h^T_{ij}$ is the metric perturbation, which includes scalar (S) and tensor (T)
perturbations. The equations of motion for the scalar  perturbations and the
fluctuations in the Q-component have been described elsewhere.  The transverse,
traceless gauge constraints and equations of motion of the gravitational wave
perturbation are
\begin{eqnarray}
&& \widetilde\nabla^i {h^T}_{ij} = \gamma^{ij} {h^T}_{ij} = 0 ,   \cr\cr
&& \Big[ {\partial^2 \over \partial t^2} + 3 H {\partial  \over \partial t }
- {1 \over a^2} \widetilde\nabla^m \widetilde\nabla_m \Big]{h^{T\,i}}_j = 0,
\end{eqnarray}
where $\widetilde\nabla_j$ denotes the covariant derivative with respect to the
spatial metric $\gamma_{ij}$. Because we have modeled the Q-matter as a scalar
field, the explicit form of the equation of motion for the tensor perturbations
is  unchanged; there are no additional inhomogeneous tensor fields or
anisotropic tensor sources (as would occur if the Q-component were modeled as a
tangled web of non-intercommuting cosmic strings,\cite{PenSperg} for example). 
Hence, the only
effect on the evolution of the gravitational wave amplitude is through the
background expansion. It is then straightforward to adopt the standard
algorithms for evolving the Boltzmann equations with tensor fluctuations for
the problem at hand.  We have modified a series of standard Boltzmann  CMB
codes based on the synchronous gauge, maintaining a fixed relationship between
the initial amplitudes of the scalar and tensor power spectra, for the
computations described below.\cite{Crit,MaandB,Sel}

\subsection{Classification of QCDM Models}

The CMB anisotropy spectrum, both scalar and tensor subcomponents, is sensitive
to the detailed time evolution of the Hubble parameter $H$ since last
scattering. In the case of the QCDM models, these subcomponents are affected by
the time evolution of the equation-of-state of the Q-component, $w \equiv
p_Q/\rho_Q$, which in turn reflects the form of the potential, $V(Q)$. We find
that for a wide range of potentials, however, there are two broad categories by
which we may classify the behavior of the models. Hence, we will focus our
attention on representative models from each of these categories.

\begin{figure}
\epsfxsize=1.5cm \epsfbox[0 400 100 500]{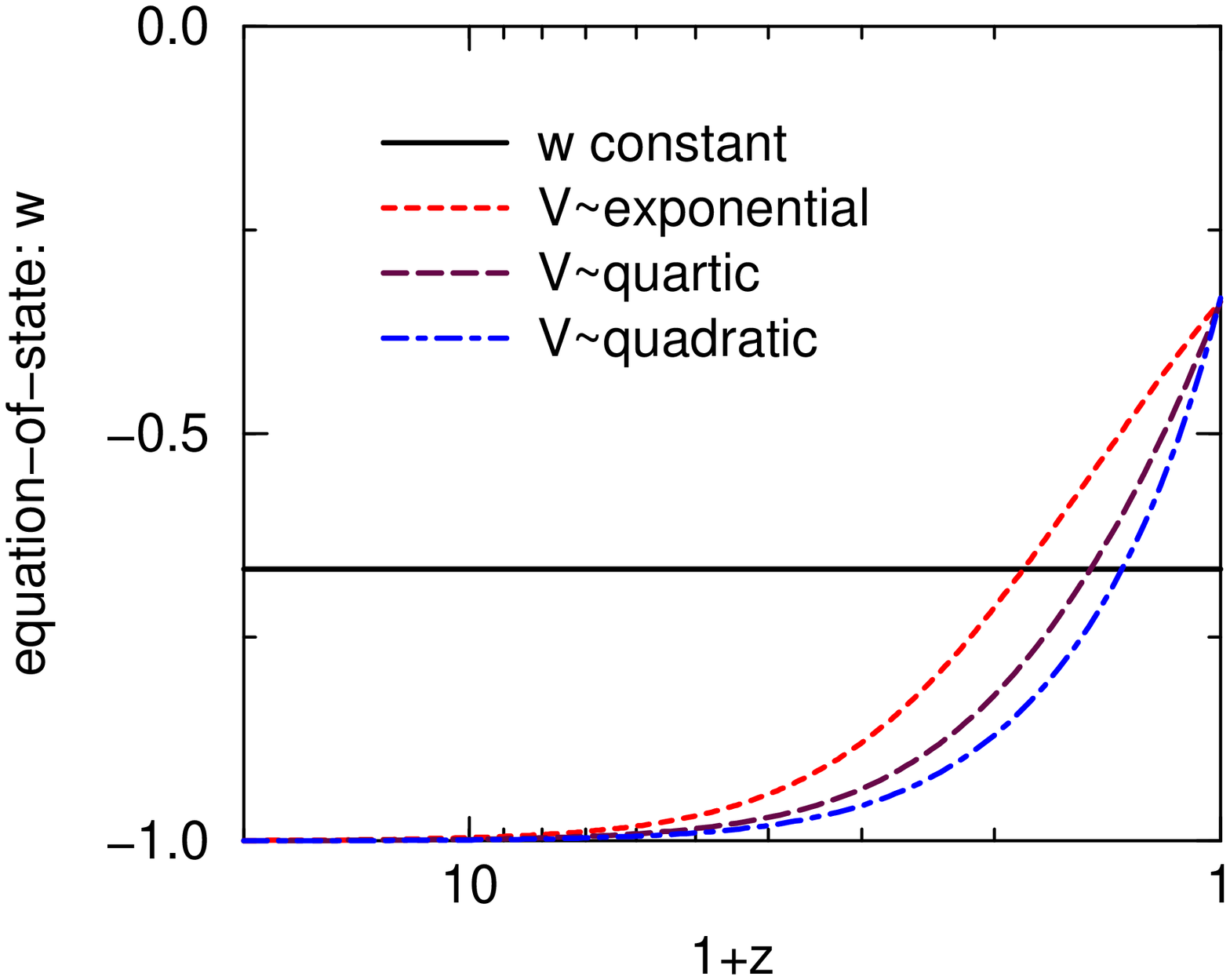}
\epsfxsize=1.5cm \epsfbox[-600 300 -500 400]{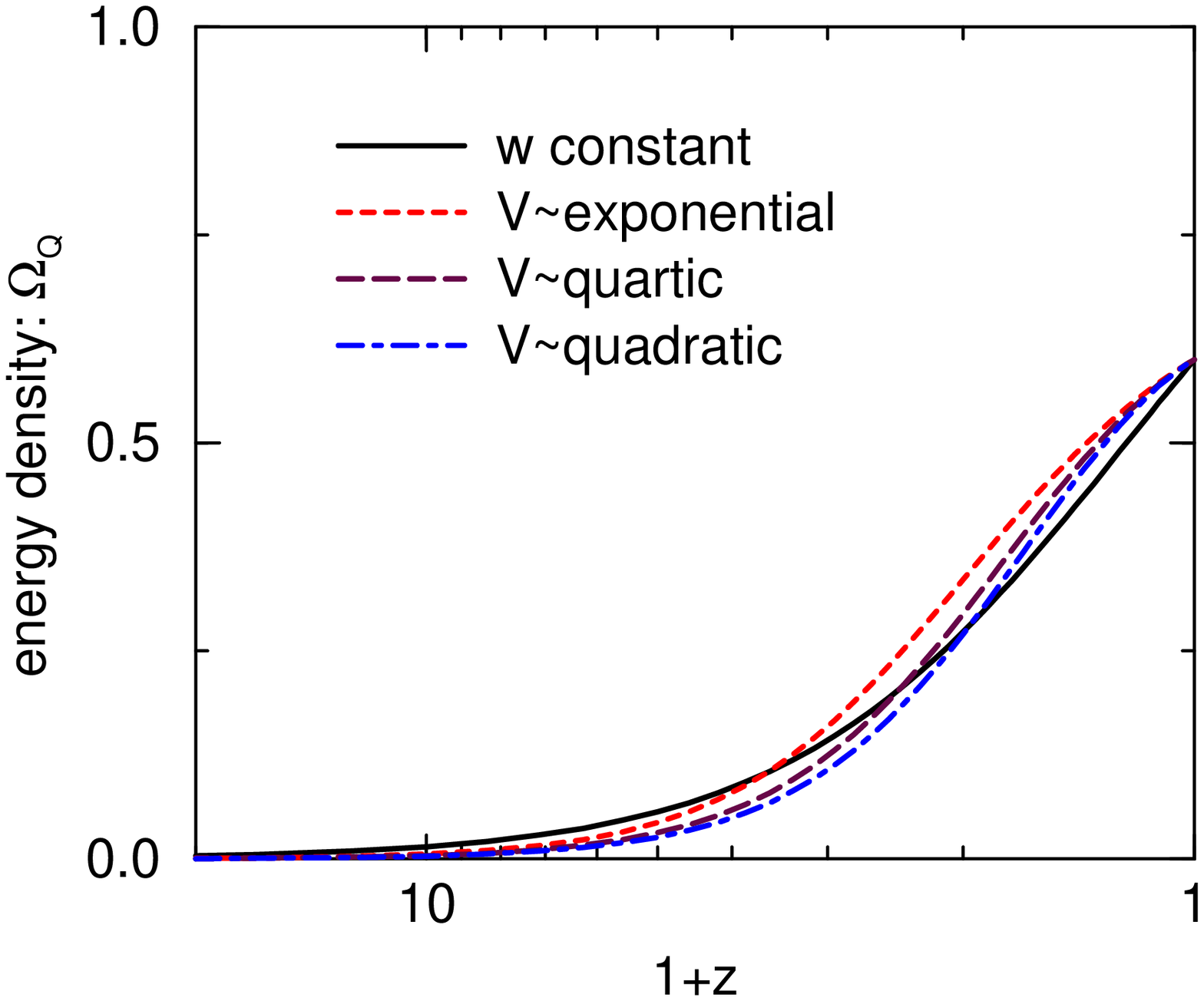}
\vskip 2.0in
\caption{For the monotonic class of QCDM models
discussed in the text, the evolution of the equation-of-state, $w$, and energy
density, $\Omega_Q$, are shown in the left and right panels. The examples
shown are  for a constant
$w=-2/3$ model and exponential, quartic, and quadratic potentials which reach
$w(t_0)=-1/3$  and $\Omega_Q=0.6$ by the present day.
}
\label{figure1}
\end{figure}

\begin{figure}
\epsfxsize=1.5cm \epsfbox[0 400 100 500]{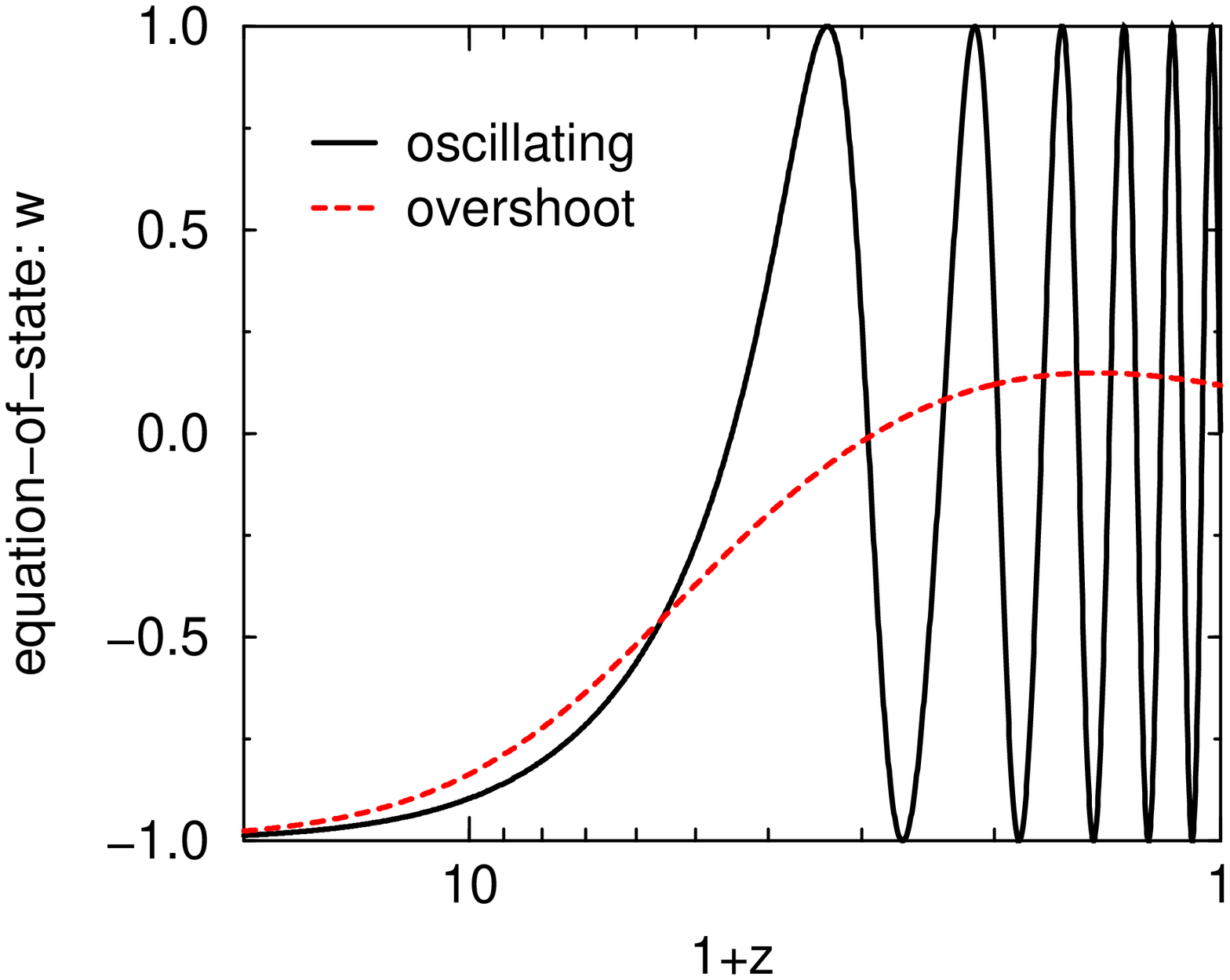}
\epsfxsize=1.5cm \epsfbox[-600 300 -500 400]{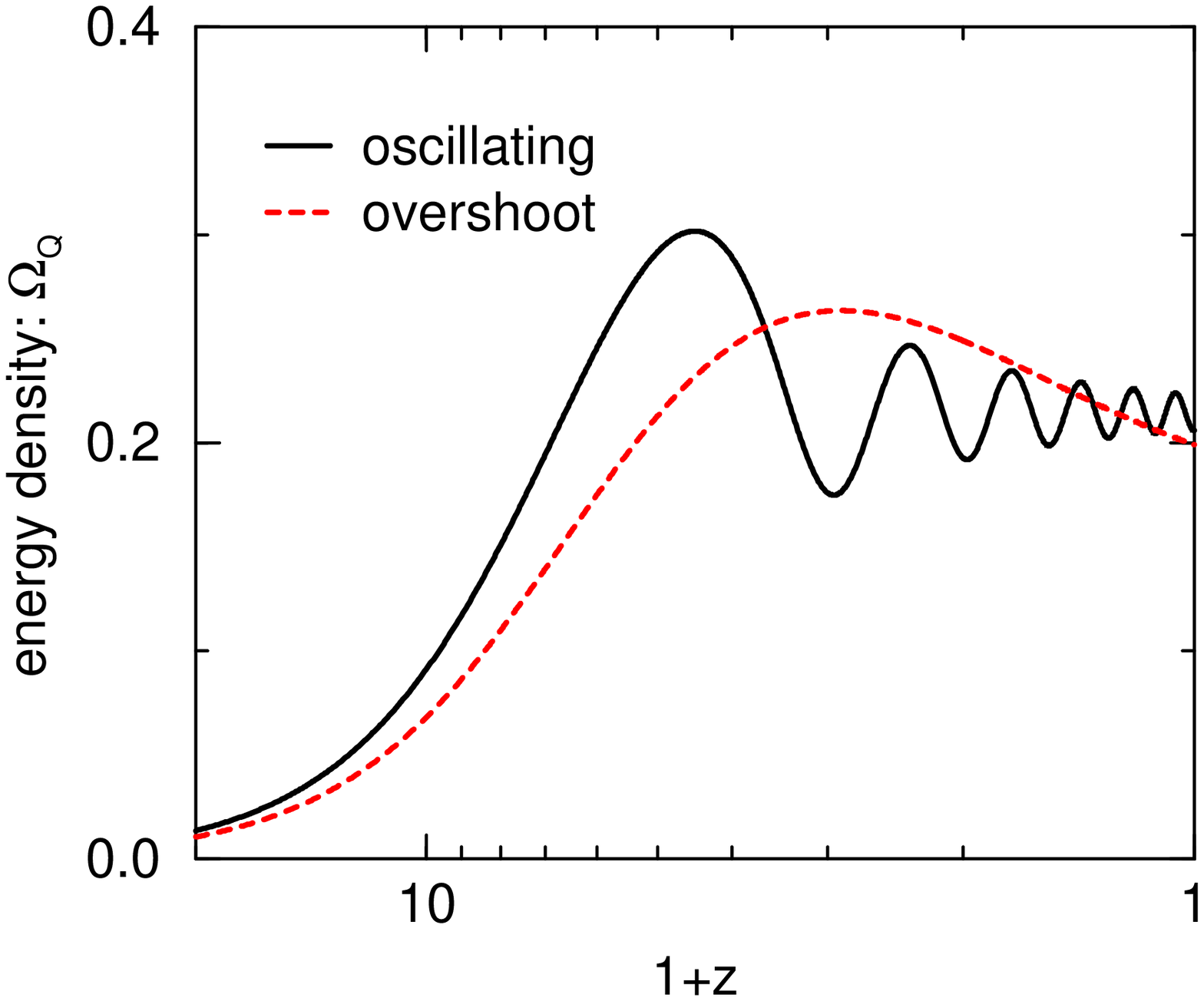}
\vskip 2.0in
\caption{For  the oscillatory class of QCDM models
discussed in the text, the evolution of the equation-of-state, $w$, and energy
density, $\Omega_Q$, are shown in the left and right panels, for oscillatory
quadratic and  overshooting exponential potentials. In each case, $\Omega_Q
\approx 0.2$ at the present day.
}
\label{figure2}
\end{figure}

The two categories correspond to cases where $w$ increases monotonically versus
cases where $w$ has begun to oscillate about an asymptotic value by the present
epoch. In either case, the initial value of the scalar field $Q$ is assumed to
be set at a particular value by some conditions ({\it e.g.}, inflation) in the
early universe. So long as $V''(Q) \ll H^2$, $Q$ remains fixed because the
Hubble red shift term ($3 H \dot{Q}$) dominates the equation-of-motion. 
Consequently, the energy density is nearly constant during the early history of
the universe so that $w\approx -1$. Once $H$ reduces sufficiently that $Q$
begins to evolve down the potential, the balance of kinetic and potential
energy in $Q$ changes and $w$ begins to grow. Depending on $V(Q)$ and the
initial conditions, $w$ may continue to grow up to the present epoch or it may
begin to oscillate around some asymptotic value. The first possibility
corresponds to the ``monotonic" class of models. 
Here we find that the imprint of $Q$ on the
CMB anisotropy and the mass power spectrum is well-approximated by the result
obtained if $w$ is held constant at roughly the  mean value during the period
when $\Omega_Q$ is non-negligible. See Figure \ref{figure1}. 
The oscillatory
class of models corresponds to cases 
where  $Q$ evolves to a point in its potential where $w$ begins to
oscillate around some asymptotic value. One example is a potential in which $Q$
evolves towards,  and then begins to oscillate about, a minimum of $V(Q)$ 
before the present epoch. Correspondingly, $w$ begins to oscillate around a
mean value determined by the shape of the potential, {\it e.g.} $w \rightarrow
0$ for a quadratic potential ($V \propto Q^2$), $w \rightarrow 1/3$ for  a
quartic potential ($V \propto Q^4$). See Figure \ref{figure2}. Another example
is an exponential potential of the form $V = m^4 \, {\rm exp}(-\beta Q)$; as
the potential energy begins to dominate the universe, the equation-of-state
first overshoots, then relaxes towards $w \rightarrow (\beta^2/24 \pi) -1$ for
$\beta \le \sqrt{48 \pi}$. We will use this exponential example  to represent
the oscillatory-$w$ category.

Note that a given potential can belong to either the monotonic or oscillatory 
class depending on parameters and
the initial conditions for $Q$.  Namely, for some choices of the
initial value of $Q$, $w$ may have begun to oscillate around its asymptotic
value by the present epoch.  For other choices, $Q$ may not have evolved so far
by the present epoch and $w$ has been increasing monotonically.  For example,
while an exponential potential is 
illustrated both in Fig. \ref{figure1} (under
monotonic) and Fig. \ref{figure2} (under oscillatory), different initial
conditions, and values of $m$ and $\beta$, have been selected in the two cases.

\subsection{Scalar and Tensor Anisotropy Spectra in QCDM Models}

We have computed the CMB anisotropy spectrum due to the scalar and tensor
subcomponents for a number of QCDM models, which we now present.

Figure \ref{figure3} shows the full CMB power spectrum predicted by inflation
for tilted sCDM and a QCDM model with a constant $w=-1/3$ equation-of-state and
$\Omega_Q=0.7$, each with $n_S=0.9$. In both panels the total spectrum is broken
down into scalar and tensor components.  It it noticeable that in the range $2
\le \ell \le 10$ the fractional contribution of the tensor spectrum to the
total power is different in the two models.

Figure \ref{figure4} indicates how the $Q$-component changes the shape of the
tensor subcomponent of the power spectrum depending on $w$ for a set of
constant equation-of-state models.  The tensor power spectra in the first panel
have been artificially normalized to $C^{(T)}_2=1$ to compare the shapes,
demonstrating that the shape is not strongly affected by the change in $w$. In
the second panel, the curves have been properly normalized with respect to
COBE.  Here we see that the main difference is in the overall amplitude; for
increasing $w$, the fractional contribution of the tensor spectrum decreases. 
In Fig. \ref{figure5}, a similar set of panels demonstrates the effect of
changes in $\Omega_Q$ on the tensor subcomponent. While the shape is not
strongly affected, we see that for increasing $\Omega_Q$, the fractional
contribution of the tensor spectrum decreases.

In Paper I, it was already noted that, even for purely scalar metric
fluctuations, the CMB power spectrum at large angular scales (low $\ell$) in
QCDM models exhibits unusual features that do not occur in sCDM or other
conventional models.  This is owing to a combination of the modification of
cosmic expansion caused by $Q$ (that is, an integrated Sachs-Wolfe effect) and
the direct effect of fluctuations in $Q$.  Adding a tensor component can add to
further features at low $\ell$. 

\section{Generalization of the Inflationary Prediction for QCDM Models}
\label{section4}

We have discussed in Section \ref{section3} how a very large class of QCDM
models can be divided into two categories: (1) models in which $w$ is constant
or monotonically increasing; and,  (2)  models in which $w$ overshoots and
then approaches or oscillates about an asymptotic value.  We have computed the
scalar and tensor components of the CMB anisotropy for representative models of
each type for a wide range of parameters.   We have used the numerical results
to obtain a revision of the inflationary relation between spectral amplitudes
and spectral indices.

Our results are expressed in terms of  an empirical relation between the scalar
spectral index $n_S$ and  the ratios $r_2 \equiv C^{(T)}_2/C^{(S)}_2$ and
$r_{10} \equiv C^{(T)}_{10}/C^{(S)}_{10}$.   The quadrupole is a conventional
choice; we have also chosen $\ell=10$ because $C_{10}$ is only weakly-dependent
on the cosmological model (compared to $C_2$) when the predicted spectra are
COBE-normalized, and so our relations can be applied more simply.  These
differences between QCDM and sCDM results have been expressed in terms of
correction factors, $f_2$ and $f_{10}$, multiplying the known  sCDM
relations:\cite{Davisetal,Oth}
\begin{eqnarray}
r_2 &\approx& \Big[ 7(1 - n_S)  \Big] 
\times f_2(\Omega_Q,n_S) \cr\cr
r_{10} &\approx& \Big[ 4.8 (1 - n_S)  \Big] 
\times f_{10}(\Omega_Q,n_S) .
\end{eqnarray}
We have not included the higher order corrections, such as those proportional
to $d n_S/d\, {\rm ln}\, k$, which are negligibly small for most
models.\cite{Wangetal} The functions $f_{2, 10}$ are defined in the following
subsections.  In all cases, $f\to 1$ as $\Omega_Q \to 0$. The dependence on
cosmological parameters $h,\,\Omega_b$ is very weak (which is why the relations are considered
to be model-independent tests of inflation) and has been ignored.

\subsection{Monotonic Evolution of $w$}

Models in which $w$ evolves monotonically leave an imprint on the CMB that is
well-approximated by the constant $w$ models in which $w$ is set to the average
value during which $\Omega_Q$ is non-negligible.  The ratios $r_2$ and $r_{10}$
for the constant equation-of-state models are shown in Fig. \ref{figure7} for
the case $n_S=0.9$. Based on plots of this sort for a range of $n_S$, we have
obtained the following correction factors for QCDM models.  For strictly
constant $w$ models, empirical fits are valid to within $10\%$ for $0\le w \le
-1$, $0.7 \le n_S \le 1$, and $0 \le \Omega_Q \le 0.7$. 
\begin{eqnarray}
f_{2}  &=& \Big[ 1 + {10 \over 9}
(1 - n_S )(2 + w) \Omega_Q^2  \Big] \,
(1 - {\Omega_Q /x})^{g_2(\Omega_Q/x) }
\cr\cr
&& g_2(y) =  -0.21 + 2.35 y - 1.03 y^2, \qquad
x =  {8 \over 5}[1 - {1 \over 2} w - {1 \over 20}(1+w)^5] 
\label{rwapproxeqn} \\
f_{10}  &=& \Big[ 1 + {1 \over 10}
(1 - n_S )((8 +  7 w) \Omega_Q^2 + 3)  \Big] \,
(1 - {\Omega_Q /x})^{g_{10}(\Omega_Q/x) }
\cr\cr
&& g_{10}(y) =  0.18 + 0.84 y^2, \qquad
x = {3 \over 4}[1 - {2 \over 3} w + {5 \over 3}w^2 - {1 \over 2}(1+w)^5]
\label{dwapproxeqn}
\end{eqnarray}
This somewhat complicated expression is needed to have a formula that
fits the dependence of all three parameters ($w$, $\Omega_Q$, and $n_S$)
over the stated range.
In the
case of $\Lambda$CDM, where $w=-1$, the function $f_2$ agrees
with the $\Lambda$CDM result.\cite{KnoxPRD}.

\subsection{Oscillatory Evolution of $w$}

Models in which $w$ overshoots and then approaches or oscillates about an
asymptotic value are well-represented by exponential potentials in which $V(Q)$
dominates the energy density of the universe by the present epoch.  The ratios
$r_2$ and $r_{10}$ for the exponential potential models are shown in Fig.
\ref{figure8}. Empirical fits to the correction factors, for the same range of
models and with the same accuracy as described in the previous subsection, are
given below.
\begin{eqnarray}
f_{2}  &=& \Big[ 1 + {10 \over 9}
(1 - n_S )(2 + w) \Omega_Q^2  \Big] \,
(1 - {\Omega_Q /x})^{g_2(\Omega_Q/x) }
\cr\cr
&& g_2(y) =  -0.21 + 2.35 y - 1.03 y^2, \qquad x = 1 - {3 \over 5 }w  
\label{rxapproxeqn} \\
f_{10}  &=& \Big[ 1 + {1 \over 10}
(1 - n_S )((8 +  7 w) \Omega_Q^2 + 3)  \Big] \,
(1 - {\Omega_Q /x})^{g_{10}(\Omega_Q/x) }
\cr\cr
&& g_{10}(y) =  0.18 + 0.84 y^2, \qquad x = 1 - {3 \over 2 }w
\label{dxapproxeqn}
\end{eqnarray}
In these expressions, $w$ represents the present value 
of the equation-of-state,
$w = w(t_0)$.  In all cases, the initial conditions correspond to  
$w \rightarrow -1$.
In the expressions above, the limit $w=-1$ corresponds to $w=-1$ 
throughout, which coincides with the standard  $\Lambda$CDM
result.\cite{KnoxPRD} 
We note that these equations do not apply to extreme cases in which the   oscillations
in the Q-matter are strong enough and begin recently enough
to leave a distinct feature at large angular scales in the CMB power spectrum 
(such as a sharp peak at low $\ell$).
Because the scalar and tensor subcomponents for
such a model depend sensitively on the detailed evolution of the Q-component,
 a correction factor must  be computed model-by-model.

\section{Conclusion}
\label{sectionend}

\begin{figure}
\epsfxsize=1.5cm \epsfbox[0 400 100 500]{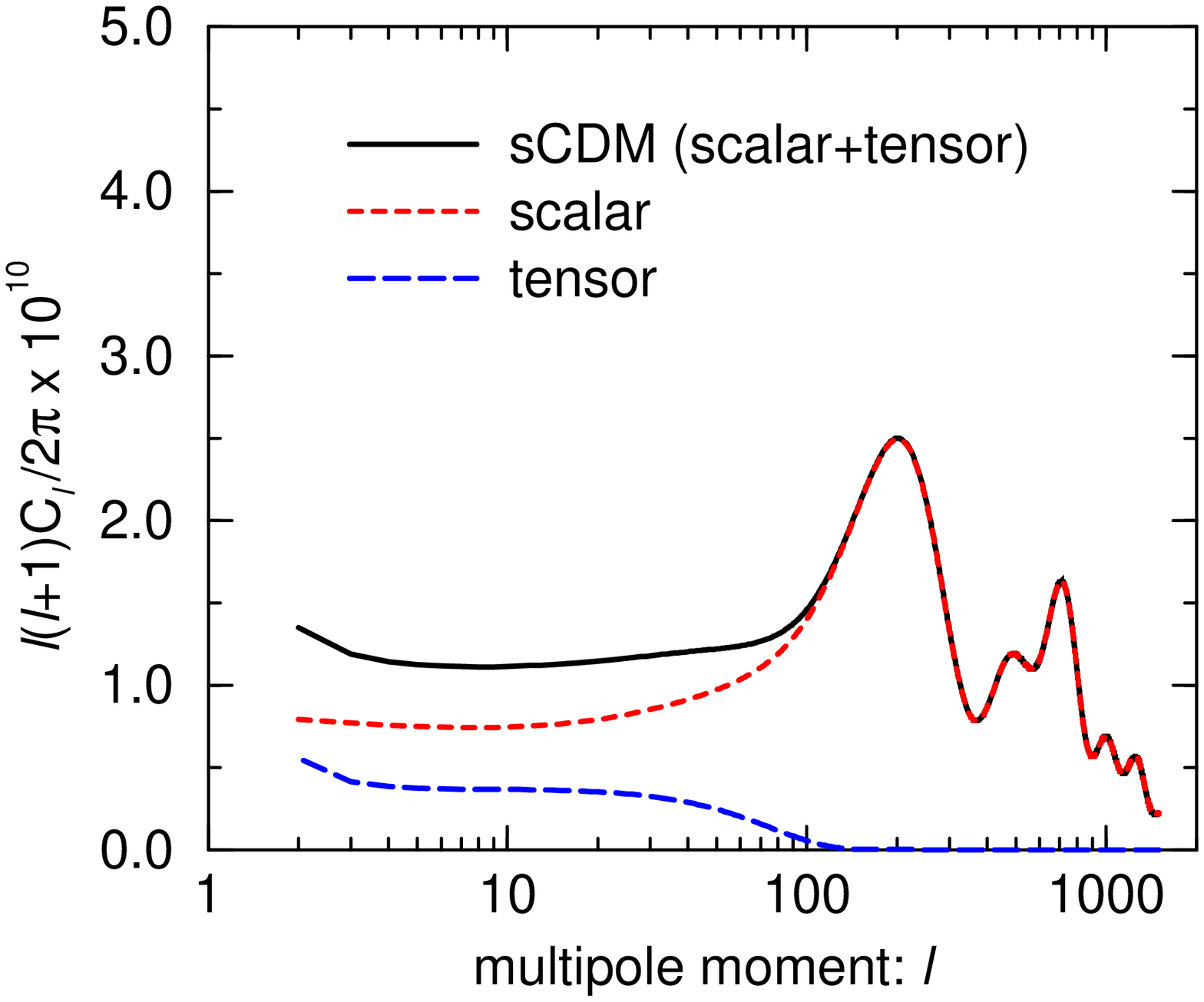}
\epsfxsize=1.5cm \epsfbox[-600 300 -500 400]{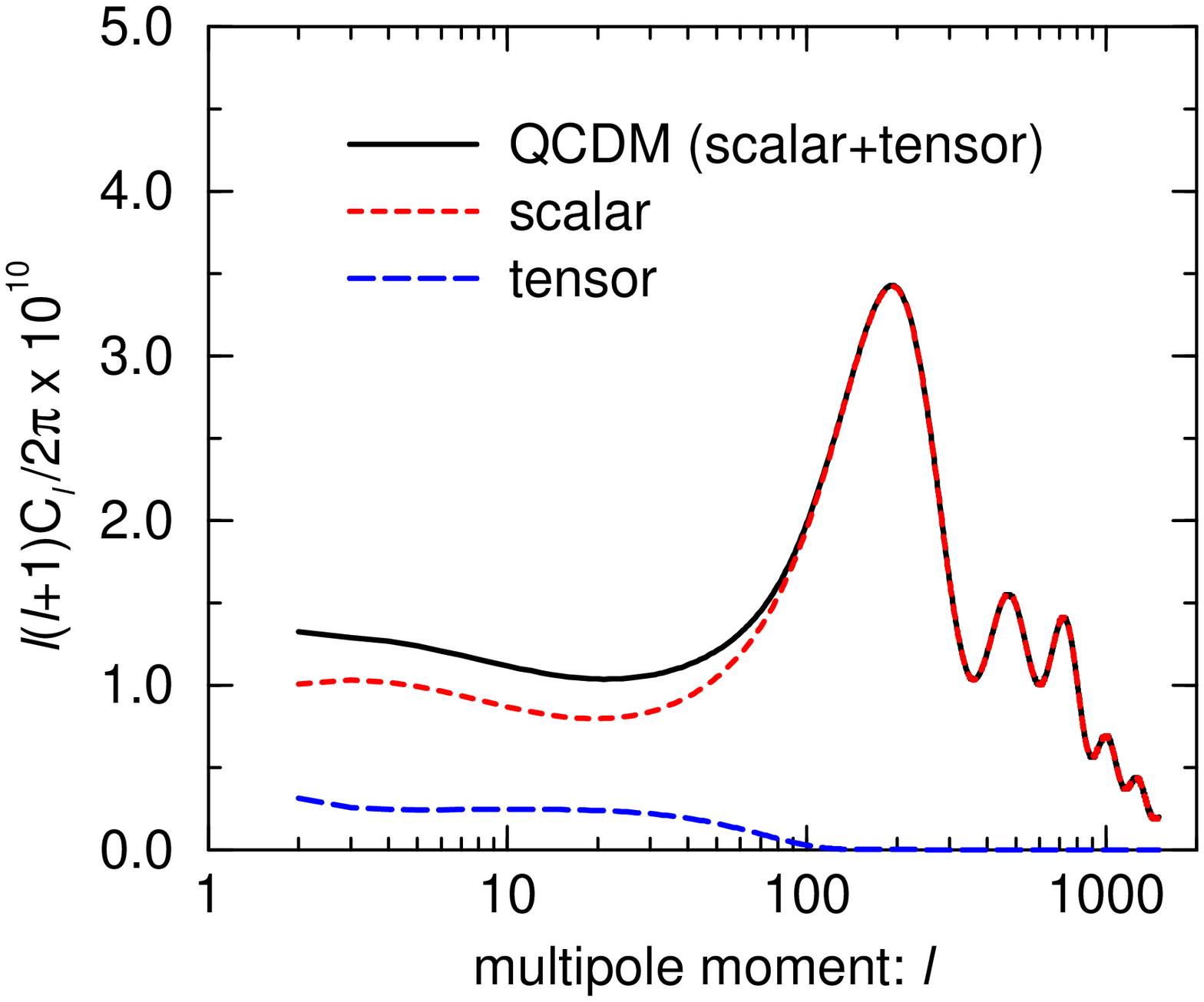}
\vskip 2.0in
\caption{The CMB anisotropy spectrum, decomposed into the scalar
and tensor subcomponents, for a tilted sCDM model
 and a tilted QCDM model with a constant
equation-of-state $w=-1/3$ and $\Omega_Q=0.7$. In each case, $n_S = 0.9$, and
the relationship between the primordial power spectra as predicted by inflation
has been maintained.  
}
\label{figure3}
\end{figure}

\begin{figure}
\epsfxsize=1.5cm \epsfbox[0 400 100 500]{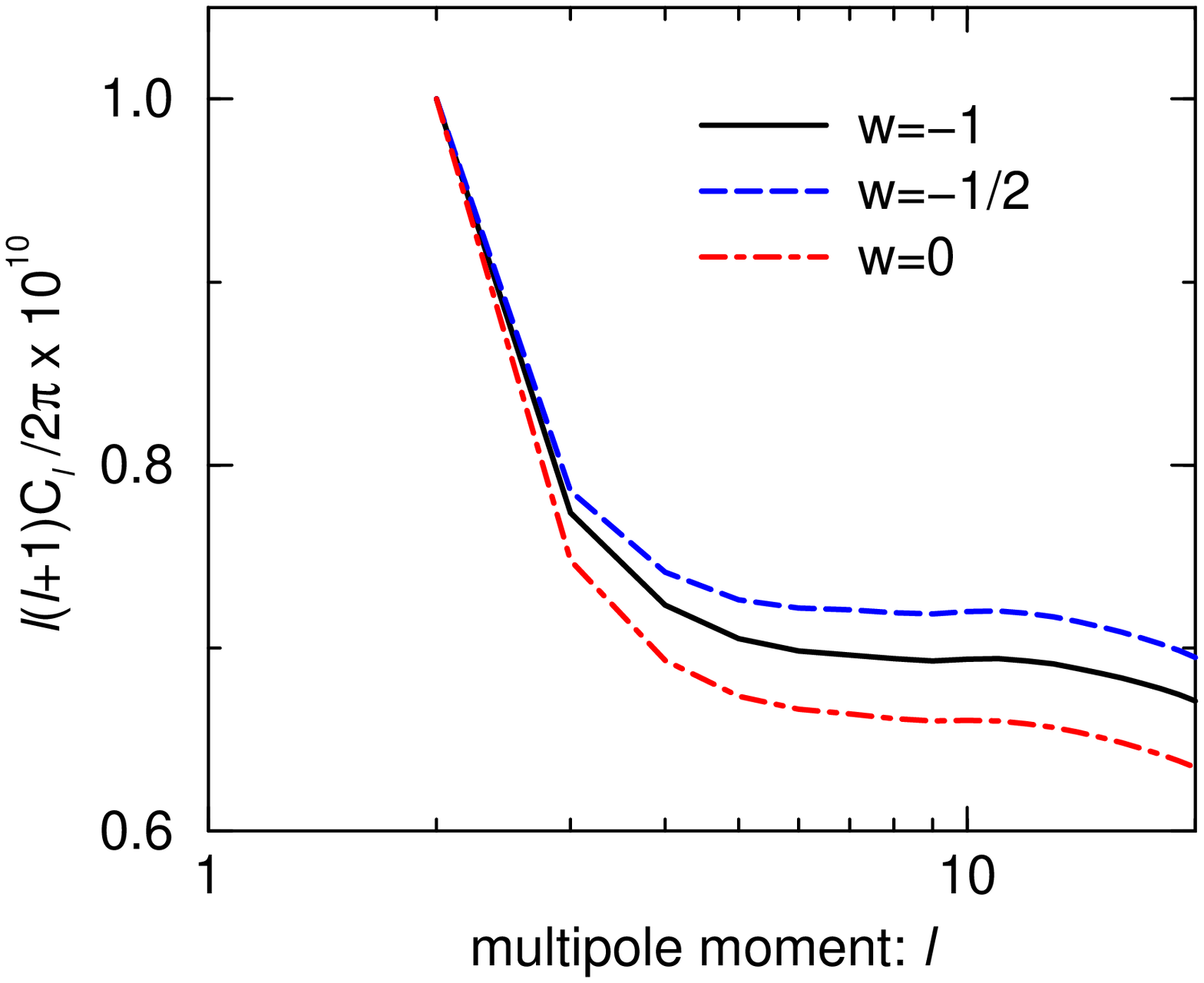}
\epsfxsize=1.5cm \epsfbox[-600 300 -500 400]{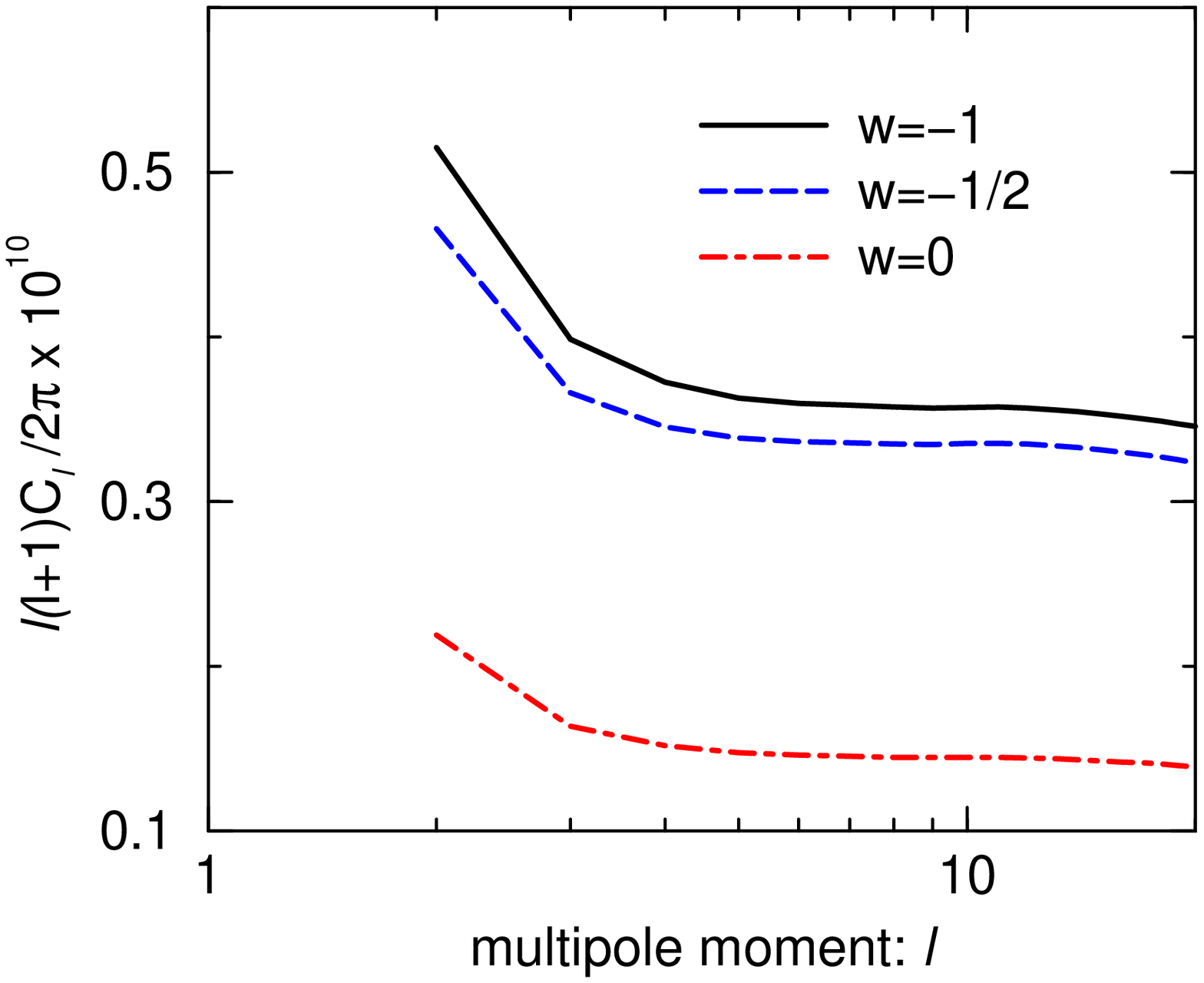}
\vskip 2.0in
\caption{The tensor subcomponent of the CMB anisotropy spectrum   
 for a series of three QCDM models with $\Omega_Q=0.5$ and $n_S=0.9$. In the
left panel, the tensor quadrupoles have been artificially set to $C_2^{(T)}=1$
in order to compare the shapes. In the right panel, the COBE-normalized
amplitudes have been restored.
}
\label{figure4}
\end{figure}

\begin{figure}
\epsfxsize=1.5cm \epsfbox[0 400 100 500]{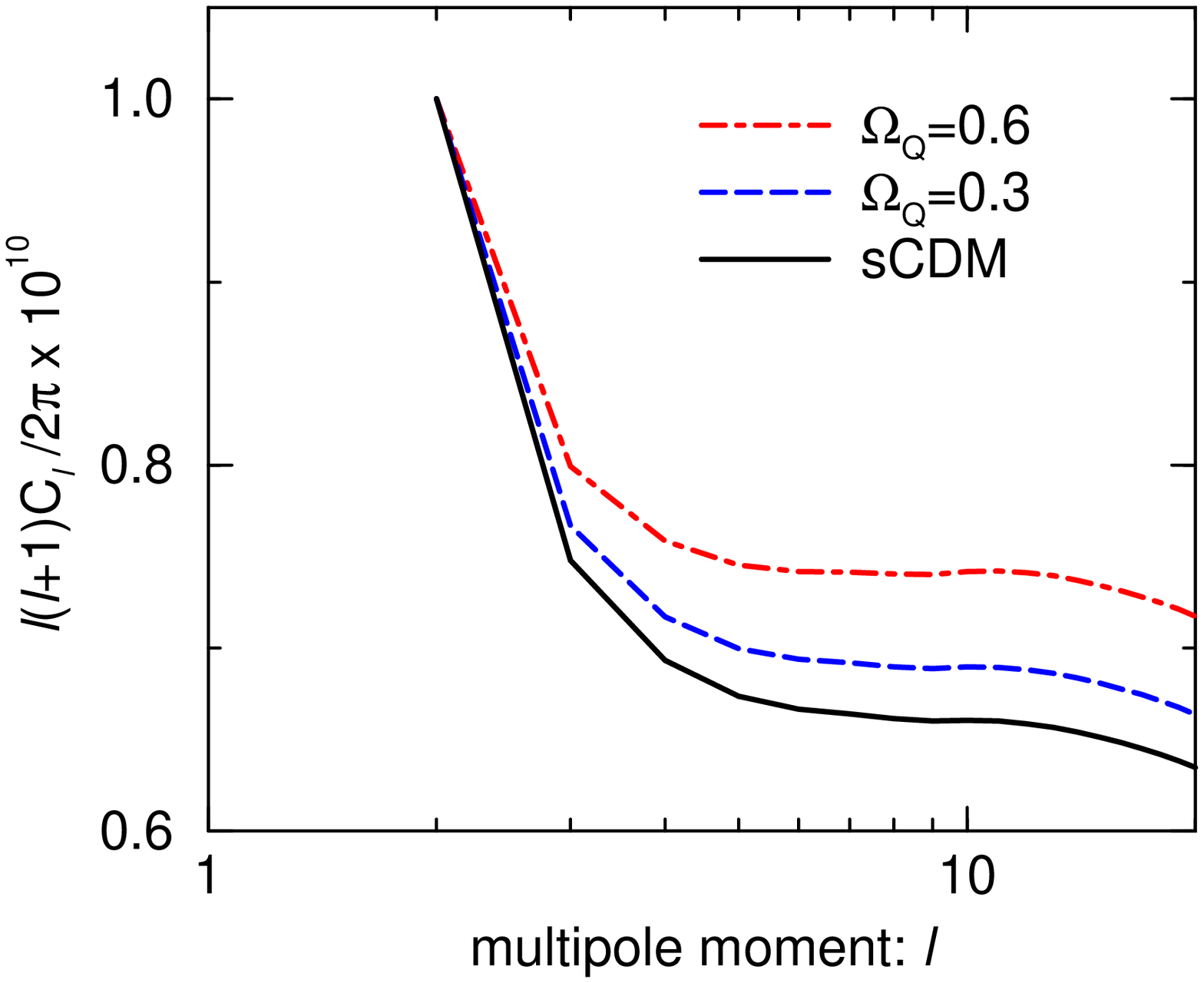}
\epsfxsize=1.5cm \epsfbox[-600 300 -500 400]{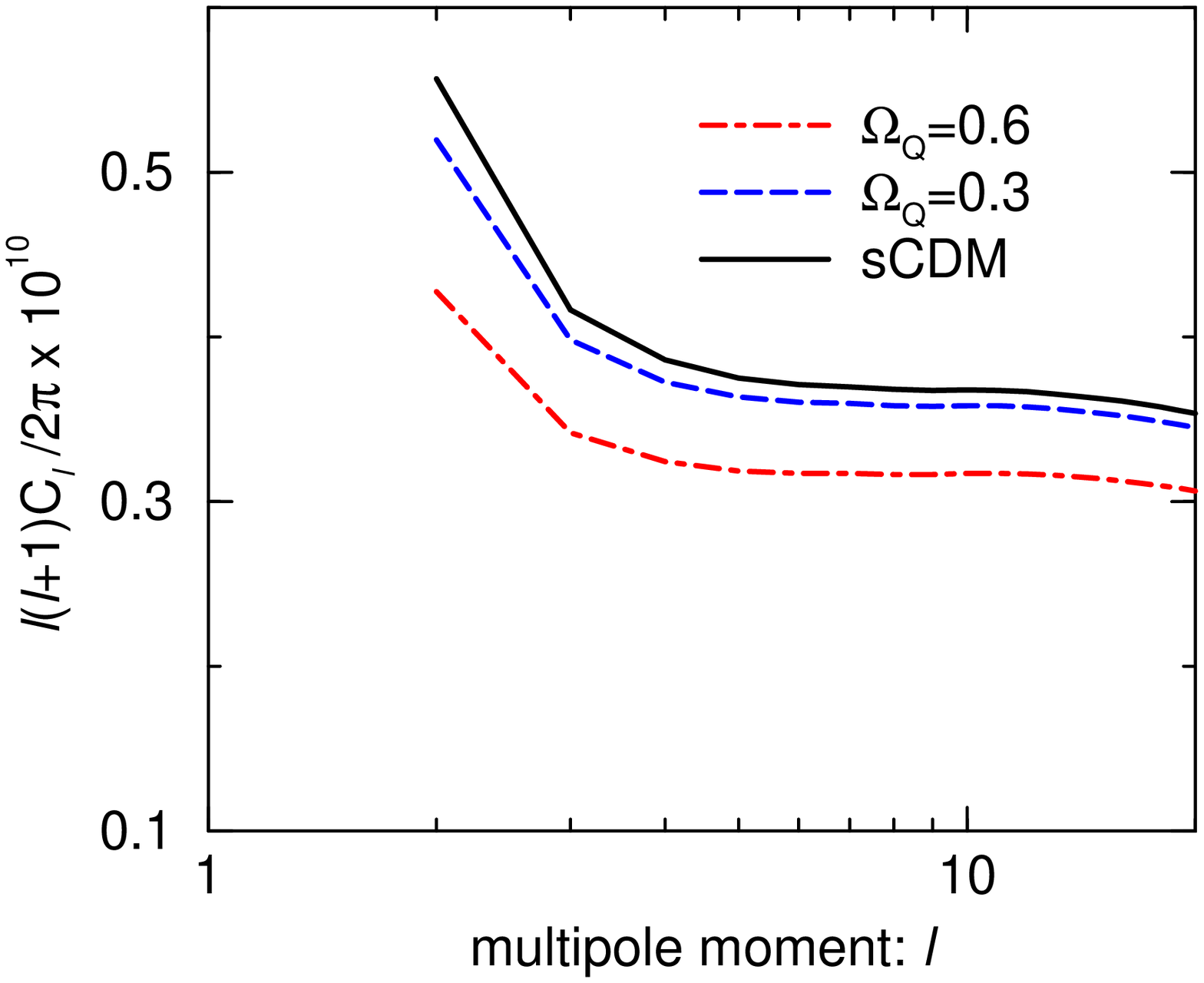}
\vskip 2.0in
\caption{The tensor subcomponent of the CMB anisotropy spectrum
for a pair of QCDM models with $w=-1/2$, and tilted sCDM, all with $n_S=0.9$.
In the left panel, the tensor quadrupoles have been artificially set to
$C_2^{(T)}=1$ in order to compare the shapes. In the right panel, the
COBE-normalized amplitudes have been restored.
}
\label{figure5}
\end{figure}

\begin{figure}
\epsfxsize=1.5cm \epsfbox[0 400 100 500]{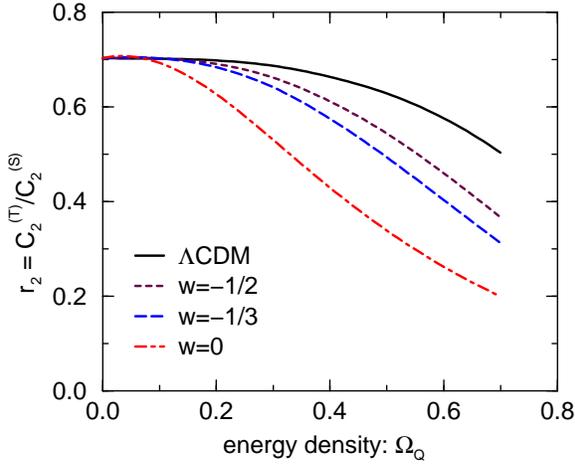}
\epsfxsize=1.5cm \epsfbox[-600 300 -500 400]{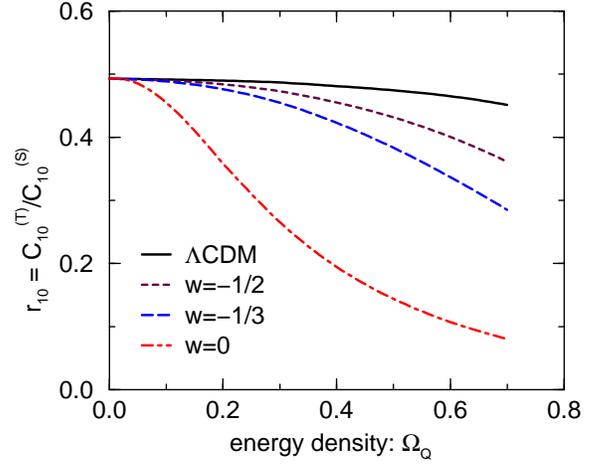}
\vskip 2.0in
\caption{The ratios $r_2$ and $r_{10}$ for a series of
 QCDM models with constant equation-of-state $w$ and  spectral
index  $n_S=0.9$, as a function of
$\Omega_Q$. 
}
\label{figure7}
\end{figure}

\begin{figure}
\epsfxsize=1.5cm \epsfbox[0 400 100 500]{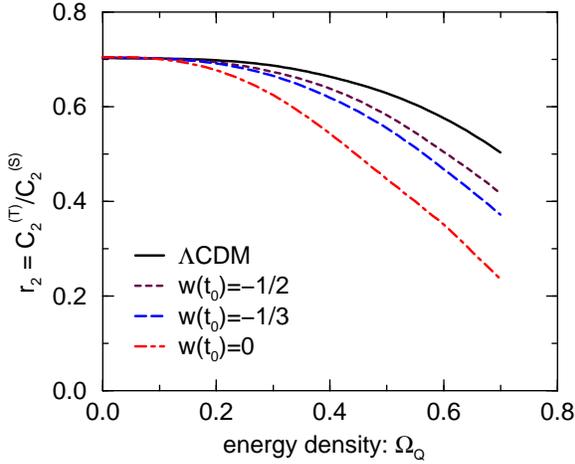}
\epsfxsize=1.5cm \epsfbox[-600 300 -500 400]{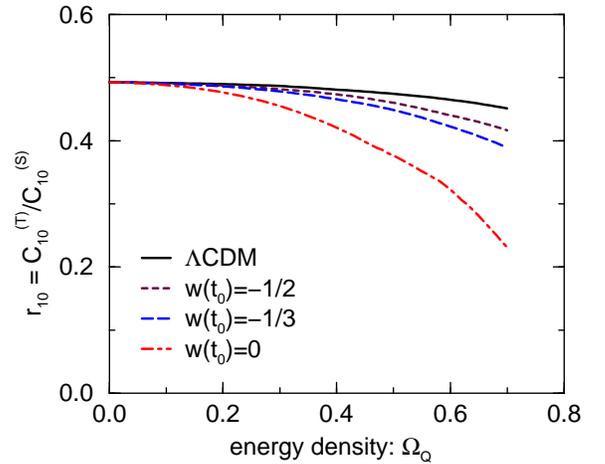}
\vskip 2.0in
\caption{The ratios $r_2$ and $r_{10}$ for a series of monotonic
exponential potential QCDM models with  spectral index
 $n_S=0.9$, as a function of $\Omega_Q$.
For these models, the equation-of-state
$w$ is time-varying  and
$w(t_0)$ is the present value of the equation-of-state. 
}
\label{figure8}
\end{figure}

We have described how Boltzmann codes to compute the CMB anisotropy power
spectrum and the mass power spectrum in QCDM models can be simply modified to
incorporate the contribution of tensor metric fluctuations, as predicted by
inflationary cosmology.  We have demonstrated that a $Q$-component has two
important effects on the tensor component. First, by modifying the expansion
history of the universe and, hence, producing an integrated Sachs-Wolfe
contribution, a $Q$-component changes the shape of the tensor anisotropy power
spectrum (see Fig. \ref{figure3}).  Secondly, since the same effect modifies
the scalar component, but by a different factor, the ratio of tensor-to-scalar
contributions to the CMB anisotropy is changed.  The net result for any given
tilt is to reduce the tensor contribution compared to sCDM when spectra are
COBE normalized (see Fig. \ref{figure4}).  Since the mass power spectrum is
normalized by the scalar contribution to the CMB anisotropy, a consequence  is
that, for a given tilt, including the tensor contribution does not reduce the
COBE normalization of the 
mass power spectrum  as much as in sCDM models. (However, as pointed out in
Paper I, the shape of the mass power spectrum in QCDM is changed and the 
small-scale power is reduced by other effects that do
not occur in sCDM models.)

Finally, the inflationary relations linking the ratio of tensor-to-scalar
multipole moments to the tilt ($n_S$) must be modified.  The most important
results presented here are the generalized relations shown in the previous
section.  These relations provide the key test for inflation in future CMB
anisotropy and polarization  measurements, now extended to include QCDM models.
Assuming inflation is correct, the relations are also used in the fitting
procedure to determine the Hubble parameter, baryon
density, and other cosmic parameters from the CMB anisotropy.\cite{Jungman} The
determination of cosmic parameters from CMB anisotropy measurements in QCDM
models, including the effect of the new relations between spectral amplitudes
and tilt, will be discussed in a forthcoming publication.

\acknowledgements

This research was supported by the Department of Energy at Penn,
DE-FG02-95ER40893.



\eject

\eject

\eject

\eject

\end{document}